\documentclass[onecolumn,11pt]{article}
\usepackage[top=1in, bottom=1in, left=1in, right=1in]{geometry}
\usepackage{amsmath,amsfonts,amscd,amssymb}
\usepackage{graphicx}
\usepackage{epstopdf}
\usepackage{overpic}
\usepackage{cancel}
\usepackage{rotating}
\usepackage{url}
\usepackage{caption}
\usepackage{color}
\usepackage{rotating}
\usepackage{multirow}
\usepackage{wrapfig}
\usepackage{mathtools}
\usepackage{subeqnarray}
\usepackage{setspace}
\usepackage{palatino} 
\usepackage[numbers,sort&compress]{natbib}
\usepackage{comment} 

\usepackage[bottom,flushmargin,hang,multiple]{footmisc}
\usepackage{lipsum}
\newcommand\blfootnote[1]{%
  \begingroup
  \renewcommand\thefootnote{}\footnote{#1}%
  \addtocounter{footnote}{-1}%
  \endgroup
}

\definecolor{header1}{cmyk}{0,0,0,1}

\newcommand{\br}{\mathbf{r}}
\newcommand{\ba}{\mathbf{a}}
\newcommand{\bu}{\mathbf{u}}
\newcommand{\bx}{\mathbf{x}}
\newcommand{\be}{\mathbf{e}}

\title{\LARGE{\vspace{-.55in}\textbf{Control of Vortex Dynamics using Invariants}}\vspace{-.175in}}

\author{\normalsize{Kartik Krishna$^{1*}$, Aditya G. Nair$^2$, Anand Krishnan$^1$, Steven L.  Brunton$^1$, Eurika Kaiser}\\
\footnotesize{$^1$ Department of Mechanical Engineering, University of Washington, Seattle, WA 98195, United States}\\
\footnotesize{$^2$ Department of Mechanical Engineering, University of Nevada, Reno, NV, 89557, United States}}
\date{}
\begin{document}
\maketitle

\blfootnote{$^*$ Corresponding author (karkris3@uw.edu).}
\vspace{-.2in}
\begin{abstract}
Vortex-dominated flows are ubiquitous in engineering, and the ability to efficiently manipulate the dynamics of these vortices has broad applications, from wake shaping to mixing enhancement.   
However, the strongly nonlinear behavior of the vortex dynamics makes this a challenging task. In this work, we investigate the control of vortex dynamics by using a change of coordinates from the Biot-Savart equations into well-known invariants, such as the Hamiltonian, linear, and angular impulses, which are Koopman eigenfunctions. We then combine the resulting model with model predictive control to generate control laws that force the vortex system using ``virtual cylinders''. The invariant model is beneficial as it provides a linear, global description of the vortex dynamics through a recently developed Koopman control scheme for conserved quantities and invariants. The use of this model has not been well studied in the literature in the context of control. In this paper, we seek to understand the effect of changing each invariant individually or multiple invariants simultaneously. We use the 4-vortex system as our primary test bed, as it is the simplest configuration that exhibits chaotic behavior. We show that by controlling to specific invariant quantities, we can modify the transition from chaotic to quasiperiodic states. Finally, we computationally demonstrate the effectiveness of invariant control on a toy example of tracer mixing in the 4-vortex system. 

\noindent\emph{Keywords-Vortex Dynamics, Mixing, Control Theory, Hamiltonian Mechanics, Chaos, 4-Vortex}
\end{abstract}

\begin{figure}[t]
    \centerline{\includegraphics[scale=.28]{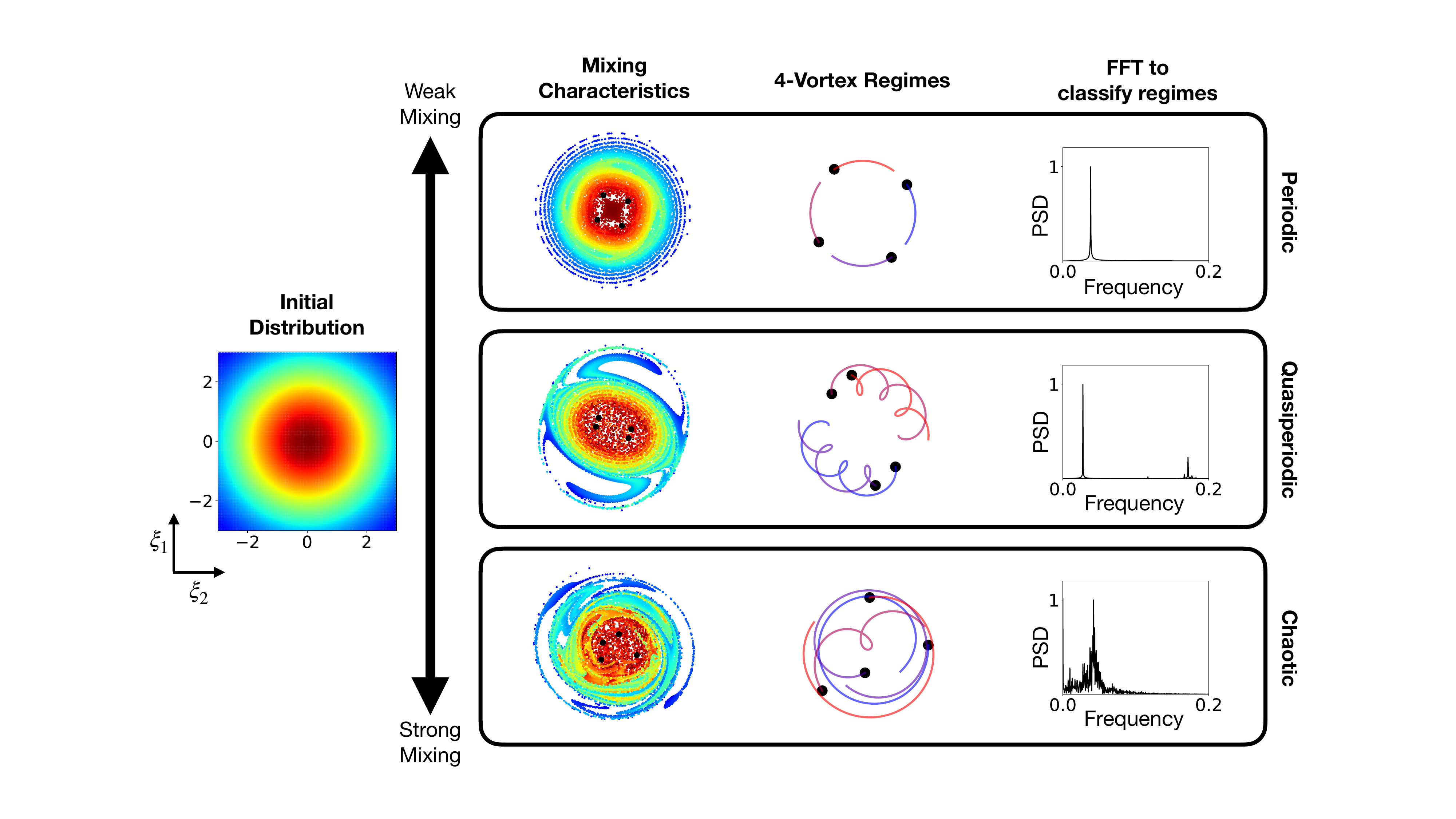}}
    \caption{{\bf Uncontrolled vortex dynamics}\rm : The key dynamic states of a 4-vortex system are highlighted: periodic (or single-frequency periodic), quasi-periodic (or multi-frequency periodic), and  chaotic (broadband frequency). These dynamic states differ by the angle the vortices initially make with the origin~\cite{boatto1999dynamics}. The control demonstrated in this work aims at switching between these different states. From left to right, we show the mixing of tracer particles (starting from the initial distribution to the left), vortex trajectories and the Fourier transform (FFT) applied to the x-component of a single vortex to characterize occurring frequencies. A broad spectrum implies chaos, whereas sharp peaks at discrete frequencies can imply quasiperiodicity
or periodicity.}
    \label{fig:baseline}
\end{figure}

\section{Introduction}
The control of vortices in fluid flows is of prime importance in many engineering applications~\cite{Dabiri2009arfm,Nawroth2012naturebio,Kinzel:2012,eldredge2019leading}. Point vortex models, governed by the Biot-Savart equations, offer a powerful framework to study vortex dynamics of ideal two-dimensional flows~\cite{newton2013n, aref2007point}.
For example, point vortex models have been used to describe various vortex phenomena in wakes~\cite{stremler1999motion, stremler2014point}.
In our work, we develop a new set of strategies for controlling point vortices based on a recently developed connection between conserved quantities and Koopman control theory~\cite{kaiser2017data,kaiser2018discovering}.

In the past, ideas from optimal control theory, such as the direct application of the Pontryagin maximum principle~\cite{noackvortex,vainchtein2006vortex} have been used to find control strategies for simple cases involving a small number of point vortices in the domain~\cite{vainchtein2002control, vainchtein2001control}. Other methods arising from chaos theory, such as OGY (Ott, Grebogi and Yorke) control~\cite{protas2008vortex} have also been used to find control strategies for periodic flows.
However, little work has been done in recasting the point vortex equations in terms of conserved quantities for control. This idea is inspired by recent work in Koopman operator theory~\cite{Mezic2005nd,Brunton2022siamreview} that shows that invariant quantities are Koopman eigenfunctions corresponding to zero eigenvalue~\cite{kaiser2018discovering}, and that it is possible to control these eigenfunctions using a Koopman with control framework~\cite{kaiser2017data}.
Direct application of model-predictive control on the governing equations \emph{has} been used to control vortex dynamics~\cite{sasaki2018model}, although, in this paper, we study the use of model-predictive control on the model formulated in terms of the conserved quantities.
This gives a global description of the dynamics written in terms of energy, angular impulse, and linear impulse, as opposed to the local description of the positions of each vortex.

We primarily focus on the four-vortex system.
The dynamics of point vortices in a domain are well understood for up to four vortices~\cite{newton2013n}.
It is known that for up to three vortices, the conserved quantities of the system can be used to reduce the dynamics to an integrable form that can exhibit mostly periodic solutions.
However, when $n=4$, the system of vortices loses integrability and behaves like a system of coupled oscillators~\cite{aref1982integrable, borisov2006}.
When the number of vortices is $n\geq 4$, the vortex system begins to exhibit chaotic behavior.
The 4-vortex system in particular, has been exhaustively studied with respect to mapping out the phase space and the period-doubling bifurcations the system undergoes as a reduced Hamiltonian energy is varied~\cite{borisov2006}.
Four vortex systems have also been studied from the perspective of mixing, where periodic, quasi-periodic, and chaotic vortex configurations have been parameterized by the angle between a single vortex and the origin~\cite{boatto1999dynamics}.
These configurations are shown in Figure~\ref{fig:baseline}.
Therefore, the four-vortex system is an ideal test case for studying invariant control due to the range of phenomena exhibited by the configuration and its simplicity.

In this work, we alter the vortex dynamics from a periodic configuration characterized by the four vortices arranged in a square (Thomson solution) as in the top panel of Figure~\ref{fig:baseline}. 
Control is applied through a virtual cylinder which is modeled as a stationary point vortex, whose circulation can be changed through model-predictive control (MPC). 
Importantly, the control is applied in a new coordinate system, defined by the vortex invariants, as this establishes a Koopman invariant subspace in which control becomes more tractable~\cite{kaiser2017data,kaiser2018discovering}. 
We first investigate changing the conserved quantities independently, with no constraints on the other conserved quantities, with relatively intuitive results:
\begin{itemize}
\item increasing/decreasing the Hamiltonian has the effect of merging/separating vortices;
\item increasing the linear momentum produces translation; and,
\item changing the angular momentum affects the variance of vorticity in the distribution.
\end{itemize}
We find that changing a single invariant typically causes other invariant values to change; therefore, we first investigate the change in a single invariant while keeping other invariants fixed through penalization in the MPC cost function.
We then show that by using multi-invariant control, we can restrict the transition to states not reachable through single invariant control alone.
We also show that conserved quantities may be used to enforce or break symmetry,  which enables both quasi-periodic and chaotic states with the same Hamiltonian/kinetic energy.
In other words, it is possible to limit the configuration to reach a quasi-periodic state as opposed to chaotic states using control.
From~\cite{borisov2006}, it is also known that Thomson solutions are surrounded by regions of quasi-periodic configurations in phase space characterized by the vortices arranged in a rectangle.
When using invariant control, we observe transitions to these nearby states, particularly, when changing the linear impulse.
Finally, we highlight the difference between single-invariant and multi-invariant control in a simple mixing example. We compare the KL divergence of these states and how much actuator energy it requires to move to them.

\section{Methodology}

\subsection{Point vortex dynamics}

Consider the governing dynamics of an inviscid, incompressible fluid given by the Euler equation as
\begin{equation}
    \frac{\partial \bu} {\partial t} + \bu \cdot \nabla{\bu} = -\nabla p,
    \label{eq:euler1}
\end{equation}
where $\bu(\bx,t)$ is the flow velocity, and $p(\bx,t)$ is the scalar pressure field at each spatial coordinate $\bx$ and time $t$.
With a systematic reduction of the symmetries of the system \citep{marsden1983coadjoint}, the vorticity field ($\omega = \nabla \times \bu$) from the Euler equations in the two-dimensional plane can be distilled to a discrete point vortex solution of the form
\begin{equation}
\omega(\bx) = \sum_{i=1}^n \frac{\kappa_i}{2\pi}\delta(\bx)
\label{eq:discretevorticity}
\end{equation}
where $\delta(\bx)$ is the Dirac delta function, and $\kappa_i$ is the strength of point vortex $i = 1, 2,\dots, n$. 
The position of each point vortex $\br_i = (x_i, y_i)$ evolves according to the Biot-Savart law as 
\begin{equation}\label{Eqn:BiotSavart}
    \frac{\mathrm{d} \br_i} {\mathrm{d}t}= \sum_{j=1, i\neq j}^{n} \frac{\kappa_j}{2 \pi} \frac{ \hat{k}  \times (\br_i - \br_j)}{\|\br_i - \br_j\|^2},
\end{equation}
where $\hat{k}$ is a unit vector normal to the $2$D plane. 
Each vortex is influenced by the induced velocity of every other point vortex in the system and the self-induced components are zero. 

An important property of the point vortex system is that it exhibits a Hamiltonian structure~\cite{newton2013n,saffman1995vortex}. 
The Hamiltonian which represents the interaction kinetic energy between the point vortices is given by
\begin{equation}\label{Eqn:Hamil}
    H = -\frac{1}{4\pi} \sum_i^n \sum_{j, i \neq j}^n \kappa_i \kappa_j \log \|\br_i - \br_j\|,
\end{equation}
such that $ {\mathrm{d} H} /{\mathrm{d}t} = 0$ for the unforced point vortex system. 
In addition to time-invariance, the Hamiltonian also exhibits invariance with respect to the translation and rotation groups~\cite{chapman1978}. 
As a consequence of Noether's theorem, the point vortex system conserves the linear and angular impulse due to translation and rotational invariance, respectively.
The impulses are 
\begin{equation}\label{Eqn:Impulses}
\text{Linear impulse}: (X,Y) = \sum_i^n \kappa_i \br_i,~~~~~\text{Angular impulse}: A = \sum_i^n \kappa_i ||\br_i||^2.
\end{equation}
By observing the form of these equations, we note that the linear impulse is directly proportional to the center of vorticity of the point vortex system, and the angular impulse is directly proportional to the variance of vorticity of the vortex configuration.
The two vortex system ($n = 2$) is trivially integrable. 
Through a series of canonical transformations~\cite{aref1979motion, aref1982integrable}, the three-vortex identical point vortex system $(n = 3)$ can be completely described by the specification of the Hamiltonian and impulses. 
This configuration has been studied exhaustively on different geometries~\cite{borisov2005sphere, borisov2005sphii}, and in itself has been useful for understanding higher dimensional vortex phenomena~\cite{stremler1999motion}.

In general, the motion of a system with four or more vortices $(n \ge 4)$ is non-integrable, exhibiting a range of dynamical behaviors including chaos. 
The invariant dynamics of the point vortex system reduces to 
\begin{equation}\label{Eq:invariant}
    \frac{d}{dt} \boldsymbol{\varphi} = 0
\end{equation}
with $ \boldsymbol{\varphi} = [H, A, X, Y]^T$. The invariant dynamics represents the global behavior of the point vortex system as opposed to the local dynamics for each point vortex of the Biot-Savart law. 
Next, we introduce forcing inputs which alter the global dynamics of the system of point vortices. 

\subsection{Multi-invariant control of point vortices}
The (forced) Biot-Savart equation can be written as
\begin{equation}\label{eq:dyn_ctrl}
    \frac{d}{dt} \mathbf{\hat{R}} = \mathbf{N}(\mathbf{\hat{R}}) + \mathbf{f},
\end{equation}
where, $\mathbf{\hat{R}} = [\br_1, \br_2, \cdots, \br_n]^T \in \mathbb{R}^{2n}$, $\mathbf{N}(\mathbf{\hat{R}})$ is the nonlinear term on the right-hand side of Eq.~(\ref{Eqn:BiotSavart}), and $\mathbf{f}$ is the external forcing added to the flow. In this work, we introduce this external forcing through ``virtual vortices''. A physical example of these virtual vortices could be the use of rotating cylinders in the domain. Rotating stirring rods are common in mixing applications. Rotating cylinders can also be used to model fans or propellers. Assuming that the vortices are sufficiently far from the stirring rods/cylinders, the viscous effects of their interaction with the vortices can be ignored. 

For a system with a large number of point vortices, multiple virtual vortices can be considered. For a system of $m$ actuator cylinders, the forcing input $\mathbf{f} = \mathbf{B}{\mathbf{u}}$ is defined such that 
\begin{equation}
    \mathbf{B}=
\begin{bmatrix}
    D(\br_1,\ba_1) && D(\br_1,\ba_2) &&
    \ldots &&
    D(\br_1,\ba_m) \\
    D(\br_2,\ba_1) && D(\br_2,\ba_2) && 
    \ldots &&
    D(\br_2,\ba_m)\\
    \vdots && \vdots && \vdots && \vdots\\
    D(\br_n,\ba_1) && D(\br_n,\ba_2) && \ldots && D(\br_n,\ba_m)
\end{bmatrix} \in \mathbb{R}^{{2n} \times {m}},
\end{equation}
where $\ba_i$ is the position of actuator cylinders in the domain. The interaction function $D$ computes the induced velocity of the control cylinders (of unit strength) on the vortices in the domain, given by the same law as the interaction between vortices as
\begin{equation}
    D(\br_i,\ba_j) = \frac{1}{2 \pi} \frac{ \hat{k}  \times (\br_i - \ba_j)}{\|\br_j - \ba_j\|^2}.
\end{equation}
The influence of the actuator cylinder on the vortices is modulated by the circulation strength of each cylinder $\mathbf{u} = [u_1, u_2, \cdots, u_m]^T \in \mathbb{R}^m$. 

We describe the global invariant dynamics of the forced system as
\begin{equation}
   \frac{\partial \boldsymbol{\varphi}}{\partial \mathbf{\hat{R}}} \cdot \frac{d}{dt} \mathbf{\hat{R}} =  \frac{\partial \boldsymbol{\varphi}}{\partial \mathbf{\hat{R}}}  \cdot \mathbf{N}(\mathbf{\hat{R}}) +  \frac{\partial \boldsymbol{\varphi}}{\partial \mathbf{\hat{R}}} \cdot \mathbf{B}{\mathbf{u}}.
\end{equation}
The first term on the right hand side of the equation cancels to zero from Eq.~(\ref{Eq:invariant}) yielding
\begin{equation}
\label{Eqn:BSInv}
    \frac{d}{dt} \mathbf{\boldsymbol{\varphi}}
    = \nabla_{\mathbf{\hat{R}}} \boldsymbol{\varphi}
    \cdot
    \mathbf{Bu}.
\end{equation}
The right-hand side of this equation is the directional derivative of the invariants with respect to the state $\hat{\mathbf{R}}$ of the vortices. We can explicitly calculate the gradient of the invariants with respect to the vortex position from Eq.~(\ref{Eqn:Hamil}) and Eq.~(\ref{Eqn:Impulses}). Due to the symplectic structure of vortex dynamics and Hamiltonian systems, the gradient of the Hamiltonian is orthogonal and equal in magnitude to $d\mathbf{R}/dt$. Thus, it can be calculated by swapping the x and y components of Eq.~ (\ref{Eqn:BiotSavart}). To compute the circulation strengths of the actuator cylinders $\bu$, we use model predictive control using Eq.~\eqref{Eqn:BSInv} as our model.

\subsection{Model predictive control}
Model predictive control (MPC)~\cite{garriga2010model,lee2011springer,mayne1997proc,morari1999model,camacho2013model,allgower2004nonlinear,eren2017jgcd} has gained widespread adoption due to its success in a range of applications, its ability to incorporate customized cost functions and constraints, and extensions to nonlinear systems.
In particular, MPC has become the {\em de-facto} standard advanced control method in process industries~\cite{qin1997proc} and has gained considerable traction in the aerospace industry due to its versatility~\cite{eren2017jgcd}.

MPC represents an optimal control problem over a receding horizon, subject to system dynamics and constraints, to determine the next control action. The optimization problem aims to solve for a sequence of control inputs $\{\bu(t_0), \bu(t_1),..., \bu(t_0 +T_H)\}$ over the time horizon $T_H$ that minimizes a pre-defined objective function $J$. The time horizon $T_H = N \Delta t$, where $N$ is the number of time steps over which the optimization is calculated, and $\Delta t$ is the discrete size of the time step of $d \tau$, given in the integral in Eq~(\ref{costfn}). Typically, only the first control input $\bu(t_0)^{\mathrm{opt}}$ is applied. The optimization problem is re-initialized each time a new measurement is collected and, thus, adaptively determines optimal control actions adjusting to model discrepancies and changing conditions in the environment and disturbances.
The most critical part of MPC is the identification of a dynamical
model that accurately and efficiently represents the system behavior in the presence of actuation.
If the model is linear, minimization of a quadratic cost functional subject to linear constraints results in a tractable convex problem.
In this work, we combine nonlinear MPC with control-affine models describing the dynamical evolution of the conserved quantities under the influence of an external control variable. 

The receding-horizon optimization problem can be stated as follows.
Nonlinear MPC~\cite{allgower2004nonlinear} aims to minimize the following quadratic objective function,

\begin{equation}
J = \int_{t_0}^{t_0 + T_H}\left[\be(\tau)^T{\mathbf Q} \be(\tau) + \bu(\tau)^T{\mathbf{ {R}}}\bu(\tau)\right]d\tau,
\label{costfn}
\end{equation}
subject to, nonlinear system dynamics,

\begin{equation}
\label{eq3}
\frac{d}{dt} \mathbf{x}(t) = \mathbf{N}(\mathbf{x}(t)) + \mathbf{B}\mathbf{u}(t),
\end{equation} 
%
%
and input constraints,
\begin{equation}
    \bu_{\text{min}} \leq \bu(t) \leq \bu_{\text{max}}.
\end{equation}
%
%
where $\bu (t)$ is the control input, $\mathbf{e} (t) \triangleq\bx (t) -\bx_{\text{goal}}$ is the difference in the desired state and predicted state along a finite-horizon trajectory.
In our work specifically, $\mathbf{e} (t) = \boldsymbol {\varphi} (t) - \boldsymbol {\varphi}_{\text{ref}}$.
The weight matrices
${\mathbf{ {R}}}\in\mathbb{R}^{q\times q}$, and
${\mathbf Q}\in\mathbb{R}^{n\times n}$
are positive semi-definite and penalize the inputs, and deviations of the predicted output along a trajectory respectively, and set their relative importance.
We define the control sequence to be solved over the receding horizon as $\{\bu(t_0), \bu(t_1),..., \bu(t_0 +T_H)\}$ given the measurement $\bx(t_0)$. The measurement ${\mathbf x(t_0)}$ is the current output of the system, whose dynamics are governed by the invariants of the  Biot-Savart equations~\eqref{Eqn:BSInv}, and is used to provide the initial condition for the optimization problem.

If the model is linear, minimization of a quadratic cost functional subject to linear constraints results in a tractable convex problem.  
Nonlinear models may yield significant improvements; however, they render MPC a nonlinear program, which can be expensive to solve, making it particularly challenging for real-time control. 
Fortunately, improvements in computing power and advanced algorithms are increasingly enabling nonlinear MPC for real-time applications.
Generally, MPC applied to the Navier-Stokes equations would result in a high-dimensional, nonlinear optimal control problem. 
By reformulating the dynamics in terms of conserved quantities, we aim to reformulate the problem as a bilinear, significantly lower-dimensional problem which is significantly more feasible.

\subsection{Specific control setup}
We set the weight matrix ${\mathbf Q}$ as
\begin{equation}
    \mathbf{Q} = 
    \begin{bmatrix}
        Q_H && 0 && 0 && 0\\
        0 && Q_A && 0 && 0\\
        0 && 0 && Q_X && 0\\
        0 && 0 && 0 && Q_Y
    \end{bmatrix}.
\end{equation}
The limits on actuation are such that
\begin{equation}
     |\mathbf{u}(t)| \leq 1.
\end{equation}

The actuator limits are set so that the maximum circulation the actuator produces does not exceed the circulation of the vortices. 
Thus, the controller will make use of the natural dynamics of the system.
The model we use for model-predictive control is given by Eq~(\ref{Eqn:BSInv}). We optimize over this model to generate an action $\mathbf{u}(t)$, which is then fed into the $\mathbf{f}$ term in Eq~(\ref{eq:dyn_ctrl}) for simulation. The time horizon is given by $N \Delta t = 0.03$ time units with time step of $\Delta t = 0.01$ and $N = 3$. The $\mathbf{\nabla_{\mathbf{\hat{R}}} \boldsymbol{\varphi}} \cdot \mathbf{B}$ matrix changes with change in state. However, for the purposes of optimization, we assume that over our short time horizon $T_H$, the state $\mathbf{\hat{R}}$ is constant. The MPCTools~\cite{rawlings2015} and CasADi~\cite{Andersson2019} packages were used for the computation of control in this paper.

\begin{figure}
    \centerline{\includegraphics[scale=.28,trim = {0 110mm 0 110mm },clip, ]{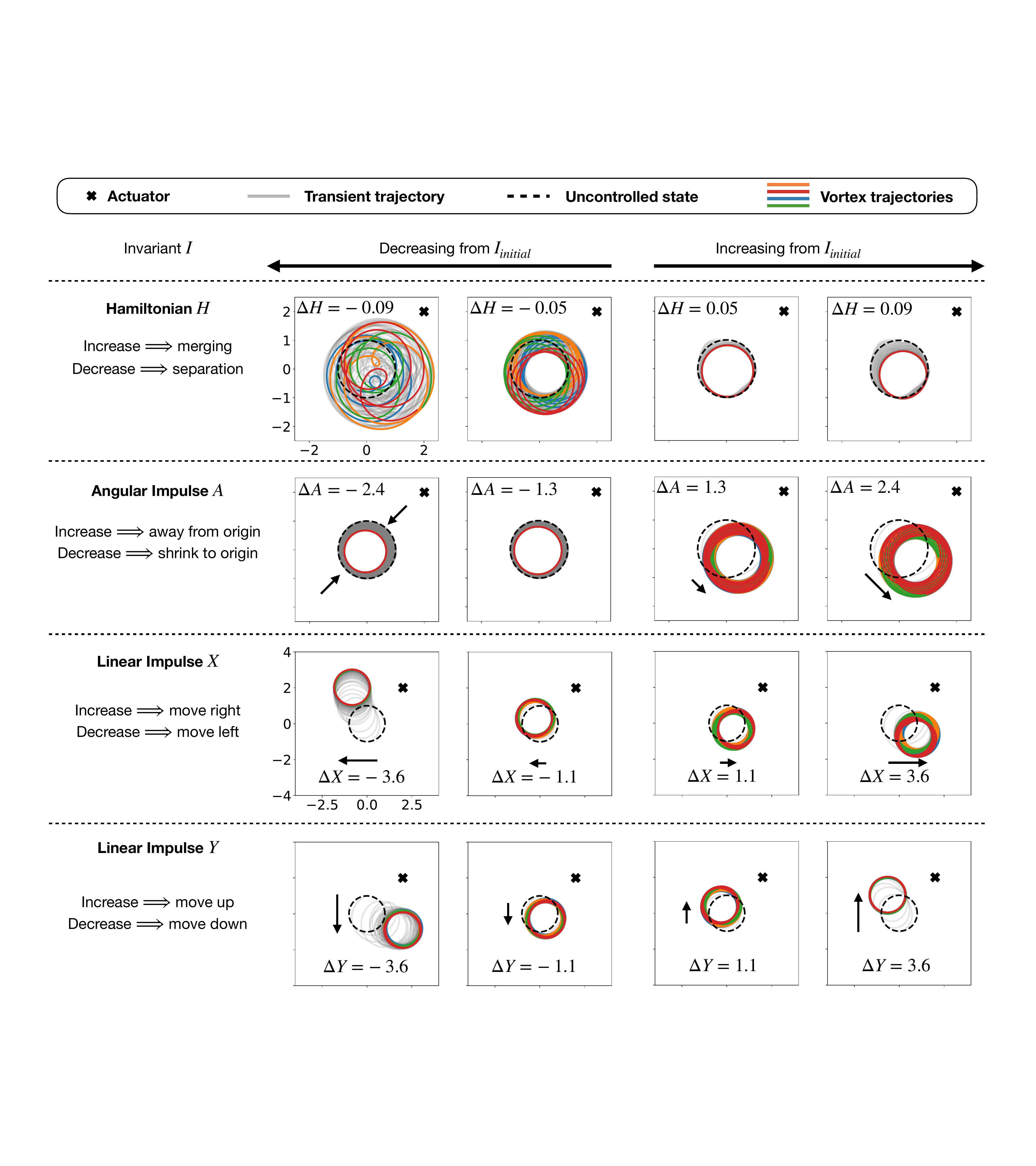}}
    \caption{Impact of reference tracking of the four invariants of the vortex system on the vortices' trajectories using an actuator in a fixed position. From top row to bottom: (1) Different dynamic regimes appear depending on a higher or lower reference value for the Hamiltonian. (2) Increasing angular impulse translates the vortices away from the origin. Negative A shows that
vortices merge closer in a similar strategy to increasing the Hamiltonian. (3-4) Change
in linear impulse manifests as translation of the vortex configuration.
}
    \label{fig:vrtx_dyn}
\end{figure}

\section{Single invariant control results}
In this section, we show how changes in invariants manifest in changes in the dynamics of vortices. 
We use the invariants to influence transitions between chaotic, quasiperiodic, and periodic states, which can in turn impact the mixing properties of the system.
For this study, we investigate the dynamics of a 4-vortex system, as 4-vortices can exhibit chaotic behaviour due to non-integrability of the governing equations. 
These configurations have also been studied extensively in the past \cite{boatto1999dynamics, borisov2006, aref1982integrable}.
The initial vortex cores are arranged in a square. The circulation $\kappa$ of all our vortices are taken to be $1.0$ for this configuration, so that the uncontrolled behaviour is for the vortices to move in a periodic circle. This is known as the Thomson equilibrium. The invariant values for this configuration are $(H, A, X, Y) = (-0.22, 4, 0, 0)$ when the vortices are all equally spaced at 1 unit from the origin.
Over the next few subsections, we explore the changes in vortex dynamics with changing invariant values, and the form of the control signal used by MPC used to change the vortex dynamics.

\subsection{Controlled vortex dynamics}
The plots corresponding to positive $\Delta H$ show that increasing the Hamiltonian has the effect of bringing vortices closer together. Decreasing the Hamiltonian has the effect of vortex cores moving apart from each other. This observation is in agreement with the form of the equation of the Hamiltonian, which is the sum of the $\log$ of the distance between each vortex pair. When there are multiple vortices present (not arranged symmetrically in a square), the MPC control law will pick one pair of vortices and bring them closer together when the Hamiltonian is increased. Using MPC and the conserved quantity framework, it is also possible to make the Hamiltonian oscillate or follow complicated trajectories. This manifests in the system as the vortices moving in closer and farther apart. The panels in the first row of Figure~\ref{fig:vrtx_dyn} show the trajectories of the 4-vortex system for different reference values for the Hamiltonian to track. We observe that decreasing the Hamiltonian makes the vortex system enter a quasi-periodic regime and further decrease makes it transition into a chaotic regime as can be seen by taking the Fourier transform of the x-component of trajectories (after it has converged to the steady state). However, increasing the Hamiltonian moves it into a periodic regime with slight translation. The mean peak in the Fourier transform plots correspond to translation of the vortex system.

The linear impulse is linked to the mean or center of vorticity for the system of vortex cores. Therefore, we observe that changing the linear impulse results in translation of the vortex configuration. The $X$ or $Y$ impulse corresponds to translations along the $x $ or $y$ axis respectively. This is because moving the vortex distribution changes the center of vorticity distribution.

Unlike the previous two cases, the vortex dynamics for angular impulse changes depending on if we are increasing or decreasing the angular impulse. For increasing angular impulse, the behavior is to translate the vorticity distribution, but decreasing angular impulse occurs by moving inwards symmetrically towards the origin.
The angular impulse of a vortex configuration is proportional to the variance from the origin. 
Thus, one way of increasing the variance from the origin is by translating the distribution away from the origin. 

A key observation across all the cases is that changing one invariant also causes a change in other invariants. For example, angular impulse is increased through translation, which, also increases the linear impulse. We can also increase the angular impulse without translation with MPC. We discuss this in a later section.

\subsection{MPC Control Law}
\begin{figure}
    \centerline{\includegraphics[scale=.28,trim = {0 75mm 0 75mm },clip, ]{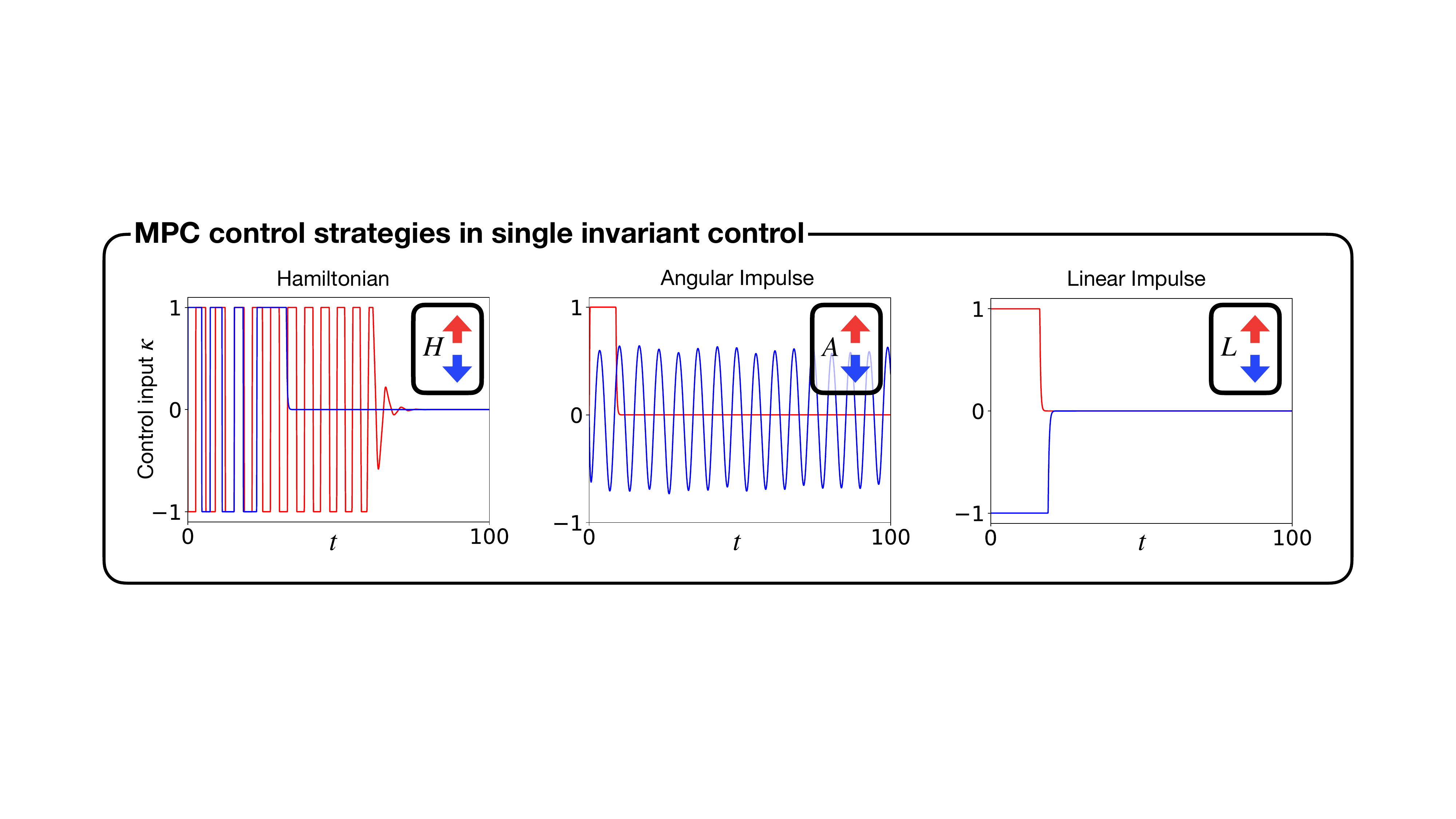}}
    \caption{MPC control signals (circulation) corresponding to the different cases of invariant control displayed in Fig.2. Red curves are associated with a higher reference value, while blue curves represent a lower reference value in the objective function. From left to right: (1) We can observe here that the red curve is oscillatory with increasing frequency in time while the blue
curve decreases the frequency in time. (2) While an increase in angular momentum results in a step function, a decrease yields consistent periodic behavior similar to the Hamiltonian but without convergence. (3) In contrast, convergence can be quickly achieved using a step function for both increasing and decreasing linear impulse.}
    \label{fig:ctrl_law}
\end{figure}
\subsubsection{Hamiltonian}
We observe in Figure~\ref{fig:ctrl_law} that the control commanded by the MPC algorithm to increase the Hamiltonian is an oscillating input, where the frequency of the oscillation increases over time. 
For decreasing, the switching frequency decreases.
In the Hamiltonian control plot, the $H$ was increased/decreased by $0.07$ units.
Vainchtein et al.~\cite{vainchtein2004optimal} showed that merging a vortex pair through control can be performed by perfectly timing a train of Dirac delta pulses in phase with the rotation of the vortex pair.
Given that increasing the Hamiltonian can be viewed as bringing vortices closer (or merging), our results are consistent with this observation, as the oscillating control is also in phase with the rotation of the four vortices.
Moreover, as the vortices move closer, the frequency of cylinder vorticity switching increases, to match the co-rotating frequency of the 4-vortex system.

\subsubsection{Linear impulse}
In Figure~\ref{fig:ctrl_law} we observe that the control law used for increasing the linear impulse takes the form of a step function. 
In the plot, the linear impulse was increased/decreased by $0.2$ units.
A step function control law causes the vortex configuration to translate along a curved trajectory as seen in Figure~\ref{fig:vrtx_dyn}. 
Changing the X linear impulse causes a translation in the x-direction and the effect is similar in the y-direction.
We observe in Figure~\ref{fig:vrtx_dyn} panels corresponding to the linear impulse that the translation also causes a net shearing effect. 
Vortices closer to the actuator translate more than vortices farther away from it because vortices impact each other proportional to $\frac{1}{r}$. 
This shearing effect is also the mechanism by which vortex trajectories move from perfectly periodic states to quasi-periodic states.
The paper by Aref et al.~\cite{aref1982integrable} diagrammatically maps out the phase space of the 4-vortex system, showing the proximity of periodic and quasi-periodic states on a KAM (Kolmogorov–Arnold–Moser) torus.

\subsubsection{Angular impulse}
\label{ai}
In Figure~\ref{fig:ctrl_law}, we observe that for increasing the angular impulse, the control law is a step function, similar to the case of linear impulse.
The cylinder actuation produces translation by ``pivoting'' the vortex system around the actuator. 
This means that the actuator can only translate up to a limit (where the vortex distribution moves diametrically across from the starting point, taking the actuator as the center). 
For the case of an actuator located at $(2,2)$, the maximum translation corresponds to an angular impulse $\Delta A \approx 130$. 
For angular impulse, the direction of translation is the direction where the control expenditure is the least, as given in the linear impulse results. 
The control law for decreasing the angular impulse is the same control law used in increasing the Hamiltonian. 
This is due to the fact that increasing the Hamiltonian causes merging, which brings the vortex distribution closer to the origin when starting from a uniform 4-vortex configuration. 
In the plots, the angular impulse was increased/decreased by $2$ units.

\section{Multi-invariant control results}
In the previous section, we discussed single-invariant control. This section is about multi-invariant control. As discussed earlier, the Biot-Savart equations for vortex dynamics has 4 conserved quantities, and we can change multiple conserved quantities at once. The advantage of simultaneously changing multiple conserved quantities is to change the distribution of vortices to states that are not possible through single invariant control. We demonstrate this with two examples. 

\begin{figure}
    \centerline{\includegraphics[scale=.28,trim = {0 30mm 0 30mm },clip, ]{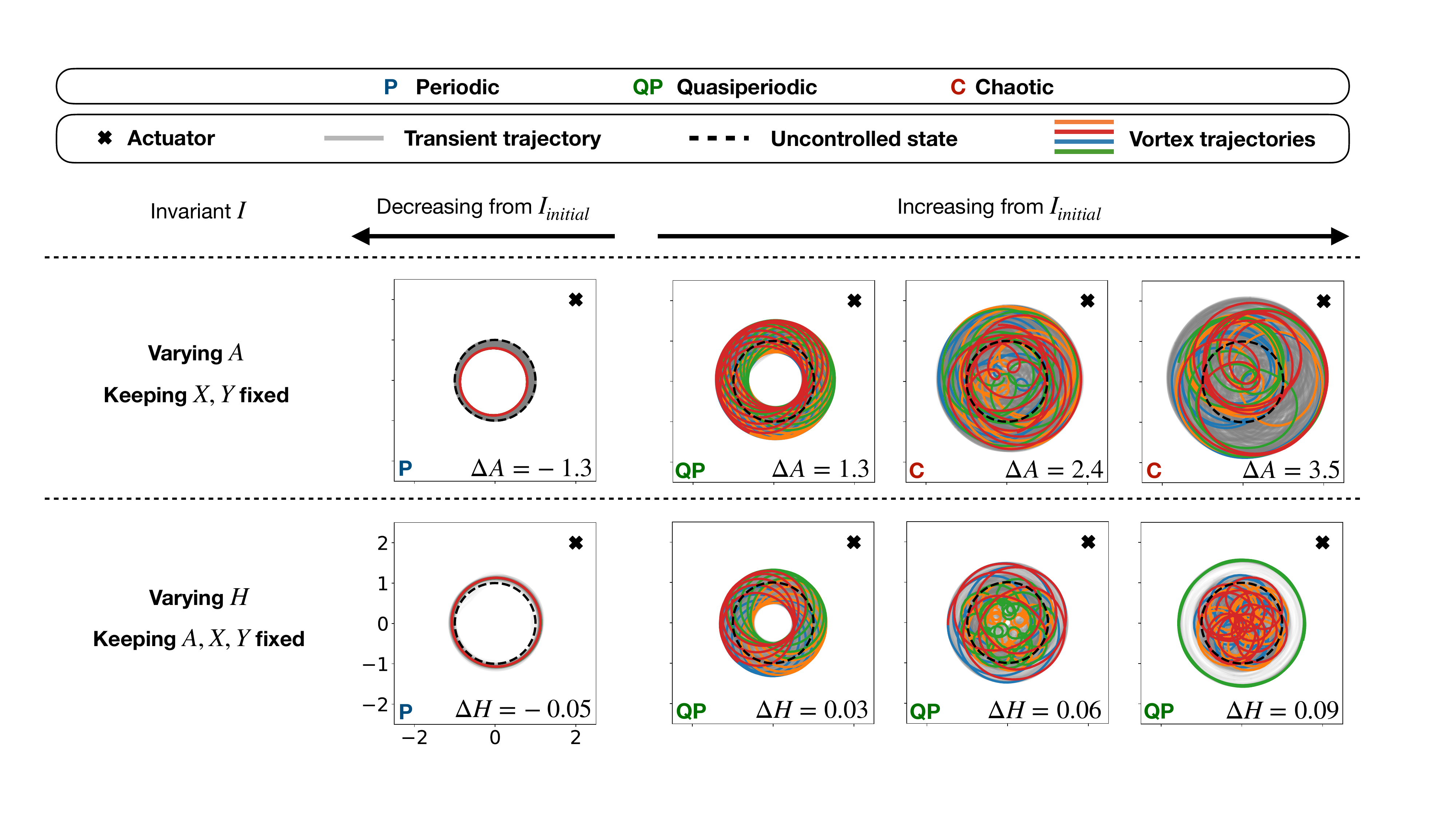}}
    \caption{This figure summarizes the resulting dynamics from multi-invariant control. The first row shows the vortex trajectories with the reference change in angular impulse. The second row is for the change in Hamiltonian}
    \label{fig:multi_inv}
\end{figure}

\begin{figure}
    \centerline{\includegraphics[scale=.245,trim = {0 30mm 0 30mm },clip, ]{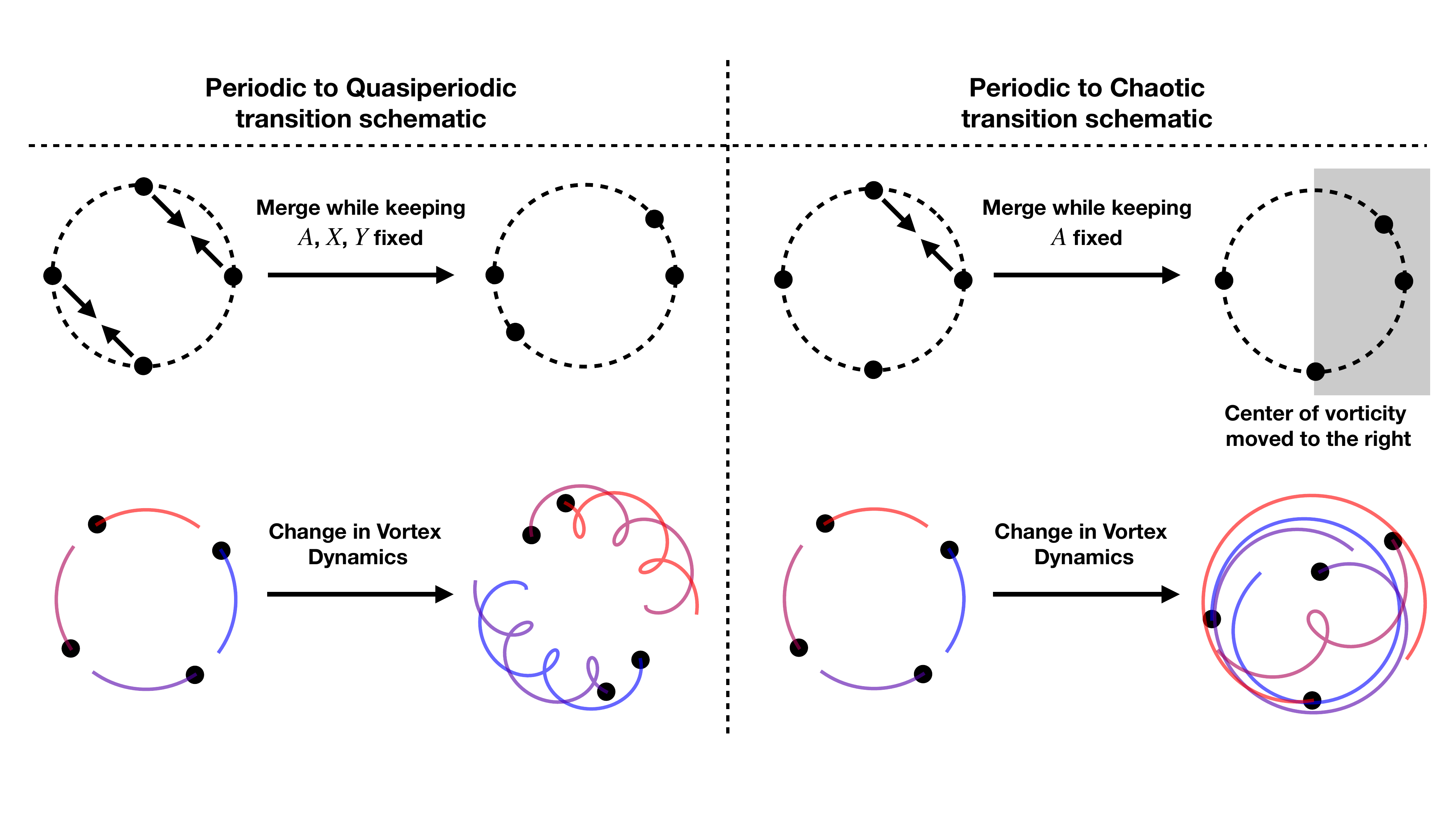}}
    \caption{{\bf Schematic diagram for transitions}\rm :  On the left, we show how multi-invariant control can be used to move between periodic and quasi-periodic regimes as seen in Figure~\ref{fig:baseline}. Here, the average radius/variance of the vortex distribution and the center of vorticity or the mean can both be controlled. When the Hamiltonian is increased under these constraints, the vortex merging is constrained to states along the dotted circle, which exhibit quasi-periodic behavior. Transitioning between different dynamic regimes is very limited using single-invariant control, whereas multi-invariant control enables one to reach a more diverse set of states. On the right, we show that a similar transition is possible from periodic to chaotic behavior by shifting the centre of vorticity to the right, thereby causing an asymmetry in the vortex configuration. These particular configurations
have been exhaustively studied independently in~\cite{boatto1999dynamics}. The key difference between these configurations is the angles the vortices make with the origin.}
    \label{fig:schem}
\end{figure}

\begin{figure}
    \centerline{\includegraphics[scale=.3,trim = {0 80mm 0 80mm },clip, ]{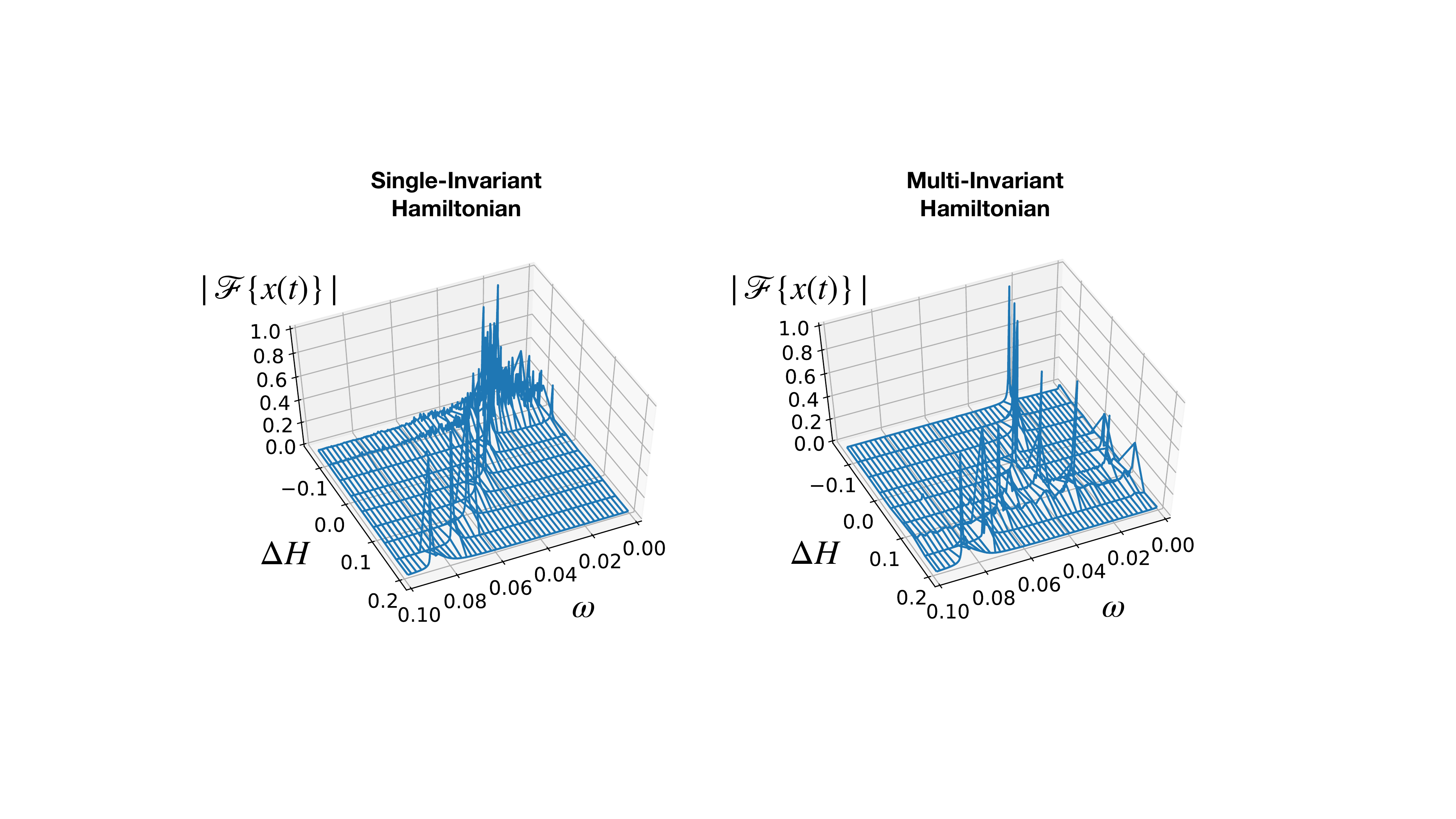}}
    \caption{This plot shows the power spectral density of the x-component of a single vortex for both single- and multi-invariant Hamiltonian control. This information can be used to classify different dynamic regimes as shown in Figure 1. We see that in single-invariant control, increasing the Hamiltonian yields periodic behavior with a single frequency, whereas decreasing the Hamiltonian modifies the dynamics towards chaotic behavior with broadband frequency characteristics.
However, for multi-invariant control, this chaotic transition does not occur. Instead, when increasing the Hamiltonian quasi-periodicity with multiple frequencies can be observed. This is due to enforcing symmetry in the case of multi-invariant control.}
    \label{fig:freq_sm}
\end{figure}

\begin{figure}
    \centerline{\includegraphics[scale=.25,trim = {0 50mm 0 30mm },clip, ]{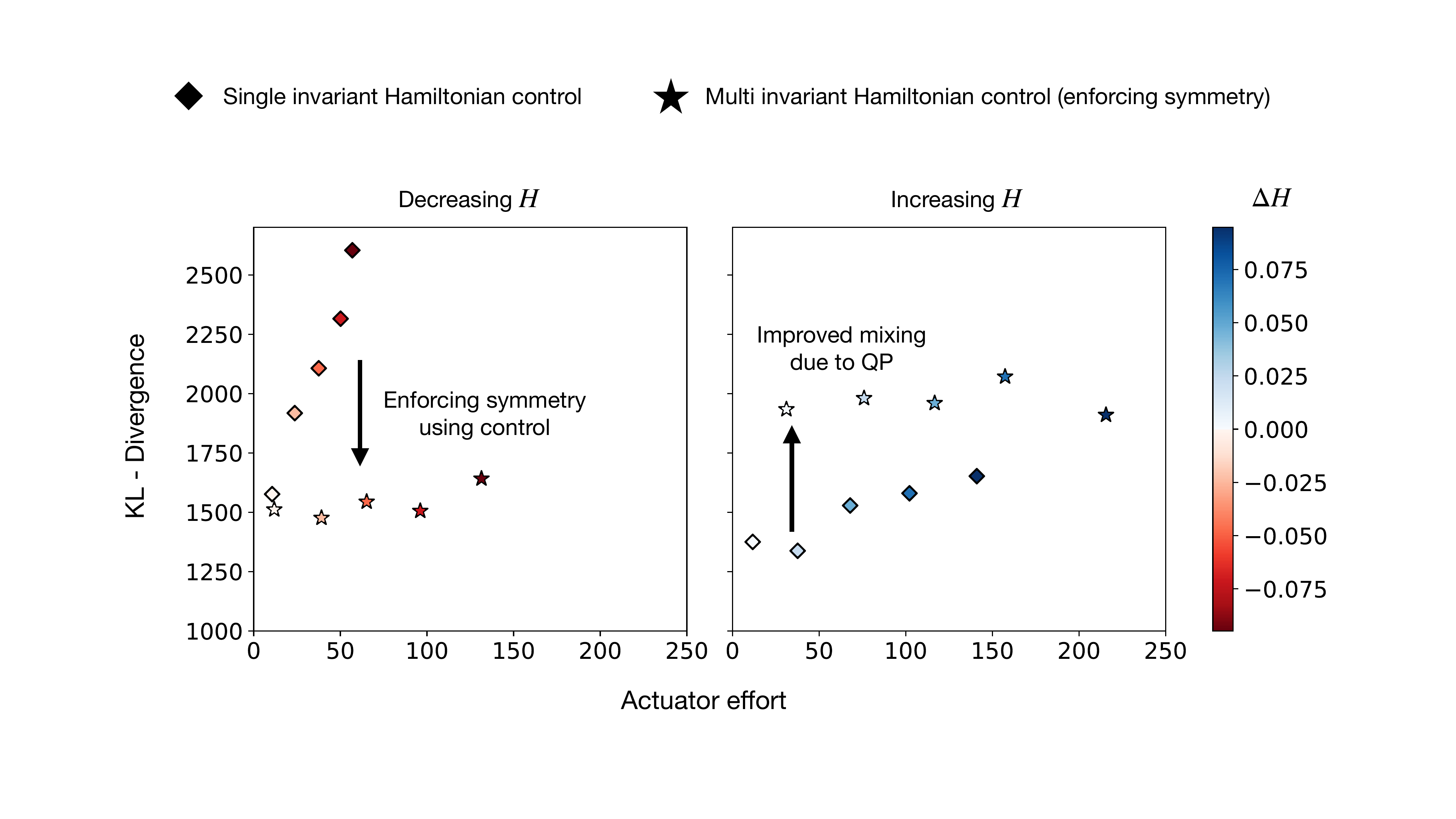}}
    \caption{This plot shows the comparison between single invariant and multi invariant Hamiltonian control in terms of mixing. The plots are scatter plots of actuator effort $\int u(t) dt$ over time on the $x$-axis and the KL-Divergence on the $y$-axis. We observe that enforcing symmetry decreases the KL-divergence in the case of decreasing Hamiltonian. This is due to vortex dynamics being restricted to periodic behavior instead of chaotic. In the case of increasing Hamiltonian, the mixing improves due to a transition to quasiperiodicity. We also observe that multi-invariant control uses more energy than single-invariant control.}
    \label{fig:Mix}
\end{figure}

\subsection{Restricting translation in angular impulse control}
We observed in \S \ref{ai} that the angular impulse increases through translation in single-invariant control. This also implies that single-invariant control changes the quantity $A$ by changing the linear impulse. We can now use multi-invariant control to change the angular impulse by penalizing the linear impulse. This forces the controller to keep the linear impulse close to $0$. The result can be seen in Figure~\ref{fig:multi_inv}, where, as we increase the angular impulse, the trajectories move outward in a manner that keeps the linear impulse close to $0$. We observe that for our choice of MPC parameters, the trajectories transition from periodic to quasiperiodic to chaotic as the angular impulse increases. Since the controller is unable to hold the linear impulse exactly at $0$, we observe chaos upon larger increases in $A$, which can be understood using the right plot schematic in Figure~\ref{fig:schem}. Moving the center of vorticity causes chaotic transitions.

\subsection{Enforcing symmetry in Hamiltonian control}
We observed earlier that using single-invariant control to change the Hamiltonian changes the other conserved quantities.
In this example, we use multi-invariant control to change the Hamiltonian, but use the MPC cost function to penalize changes in the other invariants. 
Particularly, when we increase the Hamiltonian, vortices come closer and move in a smaller periodic circle as seen in the top row of Figure~\ref{fig:vrtx_dyn}. 
However, if the other invariants are not allowed to change, we can move to quasi-periodic states as the vortices are now constrained to merge in a way that preserves the variance or center of vorticity (as shown in Figure~\ref{fig:schem}). 
We note that the final states reached in Figure~\ref{fig:multi_inv} are not exactly the final states described on the left in Figure~\ref{fig:schem}, however, Figure~\ref{fig:schem} provides a useful guide for understanding switching between regimes.
In Figure~\ref{fig:vrtx_dyn}, we observe that in the case of increasing Hamiltonian, the vortices merge such that translating the center of vorticity is prohibited, and maintaining the same average radius/variance from the center. 
We also make note that the symmetry of the configuration is generally broken when vortices move to chaotic states. 

We also perform a sweep through different increases in $H$ with and without keeping the other invariants fixed and compute the frequency spectra of the vortex time series.
Similar numerical studies have been performed in~\cite{conlisk1989} to highlight the transition between different regimes.
The power spectral density for the time series of an x-component for different baseline regimes can be seen in~Figure~\ref{fig:baseline}.
We observe that when the angular impulse and center of vorticity are fixed, transitions to chaotic states are prevented. 
We observe on the right in Figure~\ref{fig:freq_sm} that the final states are quasi-periodic when performing multi-invariant control. 
At $\Delta H = 0.09$, the 4-vortex system groups into a cluster of 3 vortices, as though it is a single vortex, and the remaining vortex orbits the 3-vortex system. 
The states of decreasing Hamiltonian converge to their reference value at an extremely slow rate, and for these cases, the simulation was ended at $80\%$ of the reference value. 
Over our simulations, we observed that states which retain the periodic nature, but reducing radius, which occur also in the case of decreasing single-invariant $A$ as seen in Figure~\ref{fig:vrtx_dyn}, take extremely long to reach the reference state and hence, use a large amount of energy.

\section{Mixing}

To enhance mixing, we are interested in rapidly dispersing some substance within the fluid such that the substance is well distributed within the domain.
There has been past work using chaotic advection through fixed stirring rods~\cite{aref1984stir, boyland_aref_stremler_2000}, and methods such as adjoint optimization~\cite{eggl2019mixing}.
In this section, we formulate the mixing problem in terms of control of conserved quantities in the 4-vortex system. 
The four vortices act as ``moving stirring rods'' for the passively advecting distribution.

For the simulations in this section, we advect a grid of particles, using a symplectic Euler scheme, for a short time through the flow field generated by the 4 vortices after the invariants converge to the reference values.
These results are plotted in Figure~\ref{fig:baseline}, where the left plot shows the initial colored tracer distribution.
Then, we compute the KL-divergence of the distribution, with respect to the initial distribution of the grid of particles to compare our results.
The KL-divergence is a measure of similarity between two probability distributions~\cite{brunton2019data}. 
We initialize a grid of uniformly distributed points and observe how the histogram of this grid distorts under the action of the flow field induced by the vortex dynamics. The right panels in Figure~\ref{fig:baseline} show the final distribution depending on the vortex configuration.
The formula of KL-divergence is given by
\begin{equation}
    KL = \sum_{x \in \chi } P(T) \log \frac{P(T)}{P(0)},
\end{equation}
where, $P(0)$ is the initial histogram distribution of tracer particles and $P(T)$ is the final distribution of tracer particles.
The single-invariant and multi-invariant Hamiltonian control cases we encountered so far encompass the wide range of dynamical phenomena that the 4-vortex system can exhibit. Hence, for the mixing study, we use this data, summarized in Figure~\ref{fig:Mix}. On the left plot, we have the results for decreasing the Hamiltonian, and on the right, we have results for increasing the Hamiltonian.
We find that in the regime of increasing Hamiltonian, for the same $\Delta{H}$, multi-invariant control can generate greater KL-divergence than in the case of single-invariant control with a slight increase in actuation effort. 
It is possible to increase the KL-divergence the most by decreasing the Hamiltonian in single-invariant control. However, this causes the vortices to move apart chaotically. In applications to mixing in confined spaces, it may be beneficial to use multi-invariant control.

\section{Discussion \& Conclusion}

In this work, we have investigated the control of vortex dynamics using vortex invariants, inspired by a recent Koopman-based control scheme.  
Specifically, the Hamiltonian and the linear and angular impulses are conserved quantities that form a set of Koopman eigenfunctions, and we demonstrate that it is possible to control these invariants to manipulate the overall vortex dynamics.  
We demonstrate this approach on the 4-vortex system, where there is a range of possible phenomena, including periodic, quasi-periodic, and chaotic dynamics.  
Using model predictive control, we show that it is possible to manipulate single invariants or multiple invariants at once, and that changing these invariants results in broad behavioral changes to the vortex dynamics (e.g., periodic to chaotic, or vice versa).  
Finally, we investigated the effects of invariant control on mixing by integrating a grid of passive tracers through the flow field generated by the 4-vortices and evaluating the KL-divergence of the distribution.

Through our study, we found that increasing (or decreasing) the Hamiltonian causes merging (or separation) of vortices. 
This is achieved by oscillating the vorticity of our control cylinder at a frequency that depends on the vortex configuration.
We also observed that changing a single conserved quantity using a vortex actuator causes a change in the other invariants. 
Multi-invariant control enables us to manipulate the dynamics of a 4-vortex system from periodic behavior to different regimes by specifying different invariant values. 
Importantly, we were able to do this by using a very short time horizon in model-predictive control. 
This is promising for the control of larger vortex systems, particularly when used with methods that cluster vortices~\cite{meena2018network, suresh2016model, taira2016network, nair2015network}.
Our method could also be potentially combined with recent methods that extract point-vortex models from flow field data~\cite{darakananda2018data}.

There are many future directions based on this work.  Extending this approach to the control of dissipative, viscous flows would be an interesting study. 
Such flow fields or vortex systems do not conserve Hamiltonian/kinetic energy, and therefore the approximate short-time conserved quantities can be learned, upon which, the framework used in this paper can be applied.
Investigating the control of higher-dimensional vortex systems would also bring this closer to relevant flows, such as turbulent shear layers~\cite{aref_siggia_1980}.
Studying lower-dimensional vortex configurations has been useful for understanding higher-dimensional phenomena, including control of 2-vortex systems for wakes~\cite{cortelezzi1996}, and control of 3-vortex systems for plasma dynamics~\cite{pentek1996stabilizing}.
This study could also be extended to understand the role of conserved quantities in modeling vortex-body interactions, as in the case of robot/animal locomotion~\cite{kelly2010self, shashikanth2002hamiltonian,vankerschaver2008geometry, wang2014high}, capturing three-dimensional fluid-structure interactions in the context of lift and drag on turbines~\cite{rodriguez2019strongly, rodriguez2020strongly}, and mitigating or deflecting large scale vortical structures in wakes behind aircraft~\cite{meunier2005physics,crouch2005airplane}.
Many of these applications involve complicated boundary conditions that require the use of additional mirror vortices.
Future work may involve formulating these objectives within the framework of varying conserved quantities.
In addition, the controllability and observability of point vortex systems can be more rigorously validated using differential geometric methods. 
Finally, the change of coordinates to conserved quantities can be incorporated into Hamiltonian/energy modeling frameworks such as port-Hamiltonian systems~\cite{mehrmann2022control, van2014port}.

\section*{Code Availability}
The code for this work has been made available on GitHub at \url{https://github.com/karkris41295/invariant-vrtx-ctrl}
\section*{Acknowledgments}
The authors acknowledge support from the National Science Foundation AI Institute in Dynamic Systems
(grant number 2112085) and from the Air Force Office of Scientific Research (AFOSR FA9550-21-1-0178).

\newpage
\bibliographystyle{plain}
 \begin{spacing}{.9}
 \small{
 \setlength{\bibsep}{6.5pt}
 \bibliographystyle{unsrt}
 \bibliography{references}
 }
 \end{spacing}

\end{document}